\documentclass[twocolumn]{aastex63}
\usepackage{lineno}
\linenumbers

\begin{document}

\title{Stellar Populations of Spectroscopically Decomposed Bulge-Disk for S0 Galaxies from the CALIFA survey}

\correspondingauthor{Mina Pak}
\email{minapak@kasi.re.kr}

\author{Mina Pak}
\affiliation{Korea Astronomy and Space Science Institute (KASI), 776 Daeduk-daero, Yuseong-gu, Daejeon 34055, Republic of Korea}

\author{Joon Hyeop Lee}
\affiliation{Korea Astronomy and Space Science Institute (KASI), 776 Daeduk-daero, Yuseong-gu, Daejeon 34055, Republic of Korea}
\affiliation{Korea University of Science and Technology (UST), 217 Gajeong-ro Yuseong-gu, Daejeon 34113, Republic of Korea}

\author{Sree Oh}
\affiliation{Research School of Astronomy and Astrophysics, Australian National University, Canberra, ACT 2611, Australia}
\affiliation{ARC Centre of Excellence for All Sky Astrophysics in 3 Dimensions (ASTRO 3D), Australia}

\author{Francesco D'Eugenio}
\affiliation{Sterrenkundig Observatorium, Universiteit Gent, Krijgslaan 281 S9, B-9000 Gent, Belgium}

\author{Matthew Colless}
\affiliation{Research School of Astronomy and Astrophysics, Australian National University, Canberra, ACT 2611, Australia}
\affiliation{ARC Centre of Excellence for All Sky Astrophysics in 3 Dimensions (ASTRO 3D), Australia}

\author{Hyunjin Jeong}
\affiliation{Korea Astronomy and Space Science Institute (KASI), 776 Daeduk-daero, Yuseong-gu, Daejeon 34055, Republic of Korea}

\author{Woong-Seob Jeong}
\affiliation{Korea Astronomy and Space Science Institute (KASI), 776 Daeduk-daero, Yuseong-gu, Daejeon 34055, Republic of Korea}
\affiliation{Korea University of Science and Technology (UST), 217 Gajeong-ro Yuseong-gu, Daejeon 34113, Republic of Korea}

\begin{abstract}

We investigate the stellar population properties of bulges and disks separately for $34$ S0s using integral field spectroscopy from the Calar Alto Legacy Integral Field Area survey. The spatially resolved stellar age and metallicity of bulge and disk components have been simultaneously estimated using the penalized pixel fitting method with photometrically defined weights for the two components. We find a tight correlation between age and metallicity for bulges, while the relation for disks has a larger scatter than that of bulges. This implies that the star formation histories of the disks are more complicated than those of the bulges. Bulges of the high-mass S0s are mostly comparable in metallicity, while bulges appear to be systematically more metal-rich than disks for the low-mass S0s. The ages of bulges and disks in the high-mass S0s appear to increase with local density. The bulge ages of the low-mass S0s also increases with local density, but such a trend is not clear in the disk ages of low-mass S0s. In addition, the age difference between bulge and disk components ($\Delta$Age) tends to increase with local density, both for the high-mass and low-mass S0s. The high-mass S0s have systematically higher $\Delta$Age than the low-mass S0s at given local density. Our results indicate that the stellar mass significantly influences the evolution of S0 galaxies, but the environment also plays an important role in determining the evolution of bulges and disks at given stellar mass.

\end{abstract}


\section{Introduction} \label{sec:intro}

Galaxies generally contain complex structural components such as bulges, stellar and gaseous disks, spiral arms, bars, and shells. Decomposing these structures is key to understanding the complex formation and evolution history of galaxies (\citealt{Kor77}; \citealt{Pen02}; \citealt{Erw15}; \citealt{Erw15b}; \citealt{Joh12}; \citealt{Joh14}; \citealt{Tab17}; \citealt{Oh20}; \citealt{Bar21}). Bulge or spheroidal components, whose surface brightness profiles roughly follow a de~Vaucouleurs law \citep{deV48}, generally contain an old stellar population with metallicities spanning the range from very metal-poor to super metal-rich. They follow the same fundamental plane as elliptical galaxies (\citealt{Ben92}; \citealt{Fal02}; \citealt{Dal18}). On the other hand, disks generally have exponential surface brightness profile \citep{Fre70} have stars with high metallicity and a wide range of ages, together with hydrogen gas, molecular clouds, dust and hot gas, heated by star formation and supernovae. The bulge component is dominantly supported by random motion with little net rotation, while the disk component is supported (and flattened) by rotation, making a flattened disk (\citealt{Cap07}; \citealt{Ems07}; \citealt{Ems11}; \citealt{Van17}). Since the two main components in a galaxy have obviously different properties, the detailed study of kinematics and stellar populations in decomposed bulges and disks is important for understanding the different processes that contributed to their formation.

Since colors can be a proxy for the stellar populations, bulge-disk decomposition has been widely applied to multi-band photometry in the past by fitting analytic functions to the galaxy light distribution in order to reconstruct the images of the various components (e.g.\ \citealt{Kor77}; \citealt{Sim98}; \citealt{Pen02}; \citealt{Erw15}). The comparison of colors between bulge and disk has shown that disks are bluer than bulges in both spirals and S0s (\citealt{Bot90}; \citealt{Pel96}; \citealt{Hud10}; \citealt{Hea14}). This implies that disky galaxies have more recent star-formation activity in the outer disk \citep{deJ96} or higher metallicities in their centers (\citealt{Bec96}; \citealt{Pom97}). However, it has also revealed that some galaxies have negative color gradients within both the bulges and disks of disk galaxies (\citealt{Ter94}; \citealt{Pel96}; \citealt{Mic00}; \citealt{Kan09}; \citealt{Hea14}), which implies the presence of older or more metal-rich stellar populations in the outer regions of these galaxies.

However, broadband colors cannot resolve the degeneracy between age and metallicity. In order to reliably measure the physical properties of a stellar population, we need to study its spectrum. Decomposition can also be done using spectroscopic data, by taking into account the different kinematic properties of bulges and disks. This involves decomposing the line-of-sight velocity distribution (LOSVD) of a galaxy into kinematic components (\citealt{Rub92}; \citealt{Kui93}). Indeed, more recently, decomposing has applied the observed spectrum into spectral components (\citealt{Coc11}; \citealt{Joh12}; \citealt{Tab17}; \citealt{Fra18b}; \citealt{Oh18}; \citealt{Men19b};\citealt{Joh21}; \citealt{Bar21}). The spectroscopic decomposition of dynamically distinct bulge and disk components is indispensable to study a detailed exploration of their individual formation histories.

A spectroscopic bulge-disk decomposition technique was introduced in \citet{Joh12} and \citet{Joh14}. They decomposed the bulge and disk components using long-slit spectra along the major axes of S0s. They fitted the one-dimensional light profiles as a function of wavelength, and then integrated these profiles to obtain the global bulge and disk spectra for each S0 galaxy. These studies found that the bulge of S0s contains systematically younger and more metal-rich stellar populations. However, they also note that using only their long-slit spectra it is impossible to determine whether these young stellar populations are genuinely associated with the bulge, or if they represent instead contamination by the central disk population.

\citet{Joh17} applied a new bulge-disk decomposition method to the Mapping Nearby Galaxies at Apache Point Observatory (MaNGA; \citealt{Bun15}) integral field spectroscopy (IFS) data. They obtained image slices at each wavelength and then constructed bulge and disk spectra from the weighted flux at each wavelength, from which they measured the stellar populations of the two components using the Lick system. \citet{Men19} also carried out a spectro-photometric decomposition for three galaxies from the Calar Alto Legacy Integral Field Area survey (CALIFA; \citealt{San12}). \citet{Tab17} introduced simultaneous spectral fitting of the two components using the Python version of the penalized pixel fitting code (pPXF; \citealt{Cap04}; \citealt{Cap17}) and applied this to three S0s from the CALIFA survey and to 302 early-type galaxies from the MaNGA survey; the populations and kinematics of the bulge and disk components have been presented in \citet{Tab19}. \citet{Bar21} also investigated the stellar populations separately in the bulges and the disks of S0s in dense environments using the Sydney-AAO Multi-object Integral field (SAMI) Galaxy Survey. These studies have found interesting constraints on S0 formation, but the consistent and systematic comparison of spectroscopically decomposed bulge and disk components between different environments has rarely been conducted, as yet.

\citet{Joh14} found that $13$ S0s in the Virgo cluster have younger and more metal-rich stellar populations in their bulges than in their disks by analyzing the Lick indices of decomposed bulges and disks from long-slit spectroscopic data. They inferred that a normal spiral galaxy hosting an old bulge and a young star-forming disk may undergo star formation quenching in its disk by gas stripping in the cluster and, simultaneously, the remaining gas may have fallen into the central bulge, triggering star formation in the bulge for the last time before the galaxy finally fades to a present-day S0. S0s in the clusters could have been through this process in their life.

\citet{Fra18b} focused on the different formation pathways as a function of stellar mass for S0s by investigating the stellar populations of $279$ S0s in the MaNGA survey from Lick indices separately measured in bulges and disks. They found that massive galaxies (M$_{\star}$ $> 10^{10}$ M$_{\odot}$) tend to have older bulges and younger disks, while less massive galaxies (M$_{\star}$ $< 10^{10}$ M$_{\odot}$) tend to have younger bulges and older disks. They argued that the role of environment is negligible. \citet{Tab19} performed spectroscopic decomposition of 302 early-type galaxies from the MaNGA survey. In their results, bulges and disks have similar ages, but bulges have systematically higher metallicities. The disks appear to have a wide range of both age and metallicity, indicating more complicated star formation histories compared to the bulges. \citet{Bar21} found that bulges are relatively redder and more metal-rich than their surrounding disks, while they did not find any notable trend with age. Their results indicate that the redder color in bulges is mainly due to the higher metallicity of the bulge stellar populations. These previous results together indicate possible differences in the bulge and disk formation of S0s between high- and low-density environments, but more consistent and systematic comparisons are necessary to clarify the situation.

Up to now two main formation pathways for S0s have been proposed: the fading of spirals to S0s (e.g. \citealt{Mor07}; \citealt{Lau10}; \citealt{Cap11}; \citealt{Kor12}; \citealt{Joh14}) and mergers or gravitational interactions (\citealt{Bek98}; \citealt{Tap17}; \citealt{Dia18}; \citealt{Riz18}). In the scenario of spirals fading to S0s, gas within spiral arms is stripped through environmental mechanisms such as ram pressure stripping \citep{Gun72}, harassment \citep{Moo99}, thermal evaporation \citep{Cow77}, and strangulation \citep{Lar80}, which results in star formation quenching. Secular evolution driven by a bar is also a possible mechanism involved in quenching of spiral galaxies, perhaps inducing a transient starburst by funnelling gas towards the centre and growing the bulge (\citealt{Com81}; \citealt{Kor04}; \citealt{Ath13}). Bars are common structures in disk galaxies in the local universe: the bar fraction has been reported to be up to $\sim 50\%$ in optical bands (\citealt{Mar07}; \citealt{Ree07}; \citealt{Bar08}) and rises to $\sim 70\%$ in near-infrared studies (\citealt{Kna00}; \citealt{Men07}). Passive spiral galaxies may be evidence for the fading of spirals to S0s. The numerical simulations from \citet{Bek02} show how cluster environmental quenching processes can transform spirals into S0s, passing through an intermediate passive spiral phase. The spiral arm structures fade over several Gyrs after the gas is stripped. Recently, however, several observational studies found that passive spirals can be found anywhere from galaxies in isolation to the centers of clusters, and hence no single mechanism can completely explain their origin (\citealt{Fra16}; \citealt{Fra18}; \citealt{Pak19}; \citealt{Pak21}; \citealt{Dee20}).

Mergers, accretion or gravitational interactions can also form S0s. The hierarchical models of galaxy formation assume that the bulges of S0s formed by a major merger of disk galaxies or by a sequence of minor merger events (\citealt{Som99}). More recent studies in \citet{Hop09} show that disks can survive even $1:1$ mergers depending on the initial gas content. This scenario seems consistent with the fact that most S0s reside in groups and with observations reporting merging relics in many S0s (\citealt{Kun02}; \citealt{Eli12}; \citealt{Bor14}). \citet{Dia18} introduced a new pathway in lower density environments, where an isolated, high-redshift compact elliptical (cE) galaxy experiences a merger with a smaller gas-rich disk galaxy. The merger remnant effectively transforms into a smooth disk without spiral structure around the cE, which ultimately evolves into an S0.

The goal of this paper is to better understand the evolution of S0s by separately relating bulge and disk stellar populations of CALIFA galaxies to local environment. While previous studies have focused either on clusters (Virgo) or field (MaNGA), this is the first study to use a range of environment consistently. Compared to MaNGA (\citealt{Hus13}), CALIFA (\citealt{San12}) offers better spatial resolution ($2.4^{\arcsec}$ in median; \citealt{Gar15}), hence better accuracy in the photometric decomposition. In addition, bulge-disk decomposition suffers from degeneracies, therefore we use the algorithm of \citet{Oh20} to overcome this issue. We use a new subroutine of pPXF for dealing with degeneracy in the solutions obtained by both photometrically and kinematically decomposing the bulge and disk components, as recently proposed by \citet{Oh20}. In particular, we focus on the systematic comparison of bulge and disk populations along a wide range of local density, which has been rarely conducted in the previous studies.

This paper is organized as follows. Section~2 describes the galaxy samples adopted from the CALIFA survey. The analysis of data is in Section~3. We present our results in Section~4. Finally, our discussion and conclusions are given in Sections~5 and~6. Throughout the paper we adopt a standard $\Lambda$CDM cosmology with $\Omega_m=0.3$, $\Omega_\Lambda=0.7$, and $H_0=70$\,km\,s$^{-1}$\,Mpc$^{-1}$.

\section{DATA and Sample Selection}  
\begin{figure*}
\centering
\includegraphics[width=17cm]{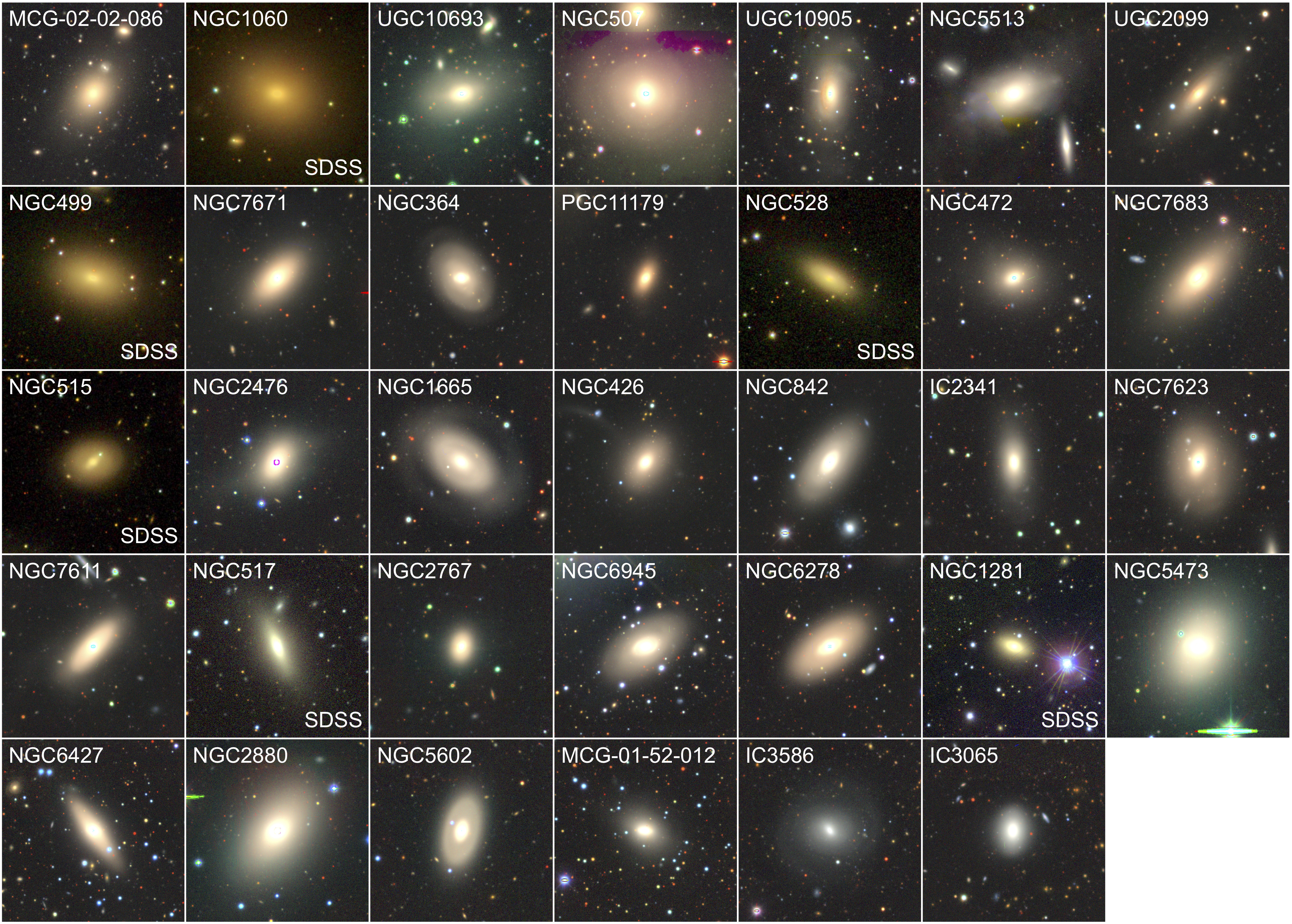}
\caption{The DECaLS and the SDSS images ($200\arcsec \times 200\arcsec$) of the 34 lenticular galaxies. The galaxies are ordered from high to low stellar mass from NASA-Sloan Atlas (NSA, \citealt{Bla11}) catalog. The name of the galaxies is shown in the top corner of each galaxy. \textbf{In case the images from SDSS, we label the `SDSS' in the bottom right corner of each image.} North is at the top and east is to the left.}
\label{F21}
\end{figure*}

The CALIFA survey (\citealt{San12}; \citealt{Hus13}) was carried out using the 3.5-m telescope of the Calar Alto observatory with the PMAS/PPAK spectrograph (\citealt{Rot05}; \citealt{Kel06}). The field of view of PPAK is $74\arcsec \times 64\arcsec$, which is filled with $382$ fibers of $2.7\arcsec$ diameter each \citep{Kel06}. The galaxies were observed with two spectroscopic setups, using the gratings V500 with a nominal resolution ($\lambda$/$\Delta \lambda$) of $850$ at $5000$ {\AA} (FWHM $\sim 6$ {\AA}) and a wavelength range from $3745$ to $7500$ {\AA}, and V1200 with a better spectral resolution of $1650$ at $4500$ {\AA} (FWHM $\sim 2.7$ {\AA}) and ranging from $3650$ to $4840$ {\AA}. More detailed information about the CALIFA sample, and the observational strategy, is available in the papers of the CALIFA team (\citealt{Wal14}; \citealt{San12}; \citealt{Hus13}; \citealt{Gar15}; and \citealt{San16}). We here analyzed galaxies using only the V500 data cube.

We visually select S0 galaxies from the CALIFA survey. Using composite color images from the Dark Energy Camera Legacy Survey (DECaLS; \citealt{Dey16}) and the 14 data release of the Sloan Digital Sky Survey (SDSS DR14; \citealt{Abo18}), we selected $87$ S0 galaxies showing a prominent spheroidal component in the center together with a disk-like structure similar to spirals, but without spiral arms. Then, we reject $25$ galaxies with heavily distorted morphology as well as paired and interacting galaxies, since galaxies with distorted features cannot be successfully constructed in the structural model from GALFIT \citep{Pen02}. We also remove two galaxies closely neighbored by bright stars, which significantly contaminate galaxy light. Finally, significantly inclined (edge-on) $26$ galaxies are also removed from the sample, because structural components in such galaxies cannot be well decomposed in our method. This process was performed with visual inspections of individual galaxies since we found that a typical disk axis ratio cut does not work for early-type edge-on galaxies with spherical stellar haloes. The final sample contains $34$ S0 galaxies (see Figure \ref{F21}). \textbf{Our S0s have stellar masses of $9.0 \leq $ log (M$\star$/M$\odot$) $< 11.5$, based on stellar masses from the NASA-Sloan Atlas (NSA, \citealt{Bla11}) catalog.} For comparison, we additionally use the nine passive spiral galaxies as described in \citet{Pak19}. 

\section{Analysis} 
\subsection{Photometric decomposition} 
We have performed a photometric bulge-disk decomposition to obtain the light fraction of both components at each resolved point in the galaxy. Firstly, the data cube has been collapsed across the whole wavelength range to produce a fair single image of the galaxy at the appropriate spatial resolution. Since the galaxies in our sample are nearby and therefore S/N of CALIFA data is fairly high, we apply GALFIT \citep{Pen02} to this collapsed image directly to fit a S\'ersic bulge and an exponential disk, adopting S\'ersic index, magnitudes, position angles, flattenings, and scale radii as free parameters, with initial estimates from \citet{Men17}. The parameters are not available for three passive spirals in \citet{Men17}. In these cases, we use the resulting values from double S\'ersic fits in one-dimension as initial estimates. We also leave the position angles and flattenings of each component as free parameters. GALFIT produces flux models of the bulge and disk components, convolved with the point spread function of the CALIFA data. Finally, we estimate the bulge-to-total (B/T) light ratio for each spaxel as the bulge model flux/(bulge model flux + disk model flux), which is used as a relative flux constraint in the spectral bulge-disk decomposition.

\subsection{Spectroscopic decomposition} 
\begin{figure*}
\centering
\includegraphics[width=15cm]{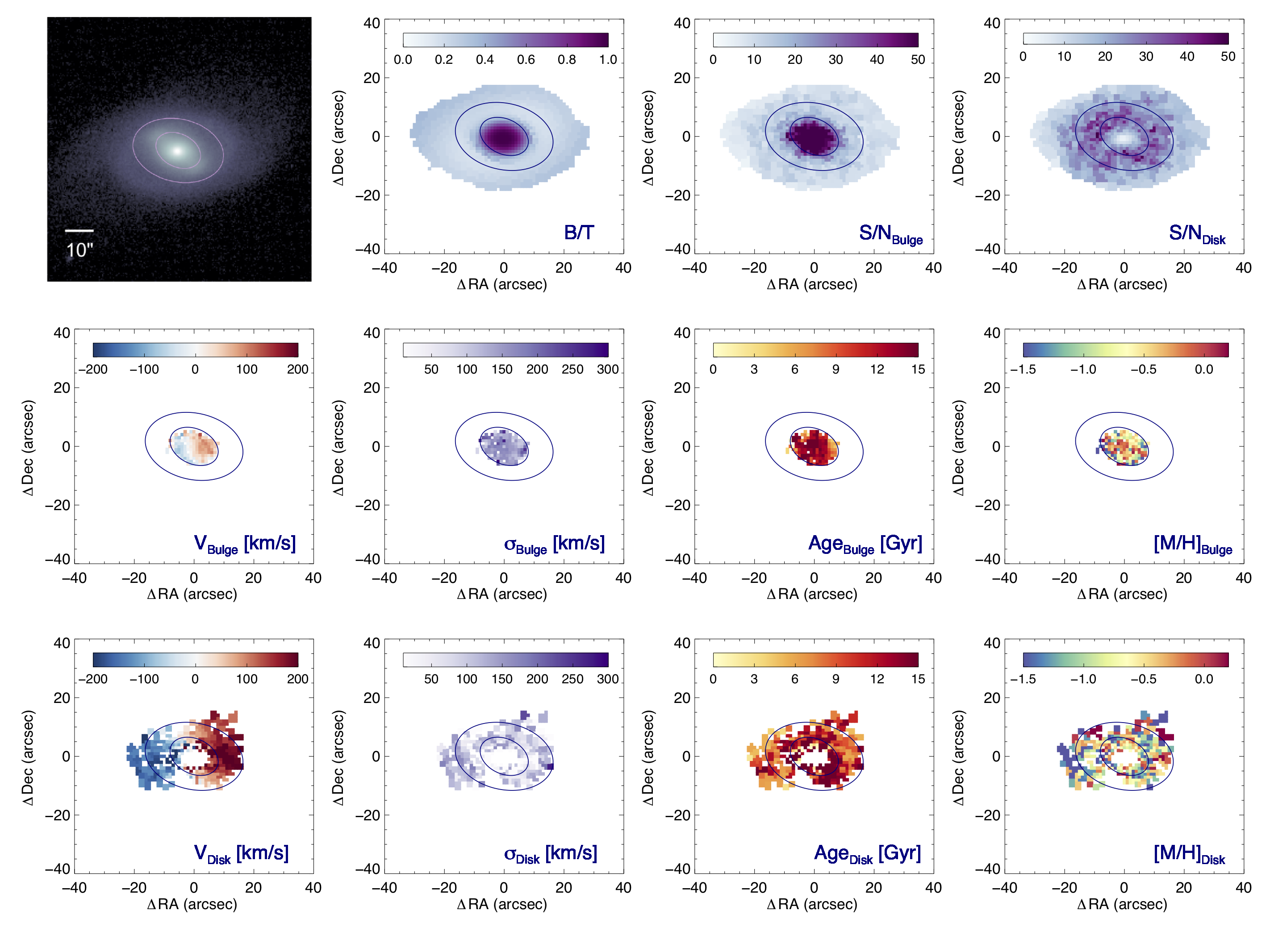}
\caption{An example of two-dimensional maps of flux, kinematics, and stellar population (UGC~1271). In the first row, the SDSS r-band image, B/T map, bulge S/N, (S/N)$_{bulge}$ = $\sqrt{B/T} \times$ (S/N)$_{continuum}$, and disk S/N, (S/N)$_{disk} = \sqrt{(1 - B/T)} \times$ (S/N)$_{continuum}$ are presented. In the middle and bottom rows, we show $V$, $\sigma$, luminosity-weighted age, and metallicity in the bulge, and the disk components, respectively, whose (S/N)$_{bulge}$ greater than $30$ and (S/N)$_{disk}$ greater than $20$. The ellipses show the 1$R_e$ and 2R$_e$ flux contours and their position angles.}
\label{F31}
\end{figure*}
We spatially bin the CALIFA data by means of the centroidal Voronoi tessellation algorithm of \citet{Cap03} using the PINGSoft \citep{Ros12} software. The minimum signal-to-noise ratio (S/N) per bin for each galaxy is set to be between 5 and 10, which are marginal values for measuring kinematics. The small S/N criterion minimizes the loss of spatial resolution in the binning result, although we may lose many spaxels when we estimate the stellar populations with stricter S/N cuts. However, because our purpose is the bulge-disk decomposition, we decided to secure spatial resolution as fine as possible rather than to bundle up low-S/N spaxels over large spatial areas, which possibly brings a negative effect on the accuracy in the decomposition. In the analysis, we used the age and metallicity averaged over the high-S/N bins for each component, which may guarantee the minimum reliability of the results.

We estimate the luminosity-weighted age and metallicity from the CALIFA data using the single stellar population model templates from \citet{Vaz10}, covering a range in ages from $0.06$ to $15.8$ Gyr and metallicities from $-1.71$ to $0.22$ (a total of $150$ templates). The templates are broadened to match the spectral resolution of the CALIFA data using the code \textsc{log\_rebin} provided with the pPXF package. As recommended by \citet{Van17}, we first run pPXF for each bin to obtain a precise noise estimate from the residual of the fit and then a second time to clip outliers using the \textsc{CLEAN} keyword in pPXF. In a third iteration, we extract the velocity and velocity dispersion using a 12th-order additive polynomial. Finally, with the extracted velocity and velocity dispersion fixed, the mean age and metallicity are estimated by using a 10th-order multiplicative (instead of additive) polynomial. We do not apply the regularization in our analysis.

We fit bulge and disk components, weighted by their relative contributions provided by the photometric decomposition at each spatial coordinate of the data cube. We use the PYTHON version of the pPXF code \citep{Cap04}, updated as described in \citet{Cap17}, which allows multiple components with different populations and kinematics to be fitted to spectra simultaneously (\citealt{Coc11}; \citealt{Joh14}; \citealt{Tab17}). The FRACTION keyword of pPXF allows simultaneous estimation of two components (i.e.\ a bulge and a disk) with a constraint on the relative weights of the two components:
\begin{equation}
f_{bulge} = \frac{\Sigma w_{bulge}}{\Sigma w_{bulge} + \Sigma w_{disk}}, 
\end{equation}
where $w_{bulge}$ and $w_{disk}$ are the weights assigned to the set of spectral templates for fitting the bulge and disk, respectively. The linear least-squares sub-problem with an exact linear equality constraint (equation 1) can be simplified by adding the following extra constraint requiring only minimal changes to the pPXF algorithm:
\begin{equation}
(f_{bulge} - 1) \Sigma w_{bulge} + f_{bulge}\Sigma w_{disk} \leq \Delta,
\end{equation}
where $\Delta$ regulates the precision required to satisfy inequality (2) and is set to a small number (we use the single precision limit $\Delta = 10^{-9}$). Both the spectral templates and the galaxy spectrum have been normalized to have a mean flux of order unity to satisfy the equality constraint to sufficient numerical accuracy.

The pPXF method can result in unphysical solutions, which is a known issue of simultaneously fitting multiple components. For example, the fit may give a bulge rotation velocity larger than the disk rotation velocity and a bulge velocity dispersion smaller than the disk velocity dispersion. Such cases are obviously unrealistic given our understanding of galaxy structure and we therefore adopt the method introduced in \citet{Oh20} to deal with them, using a `swapping' subroutine to overcome the degeneracy resulting from two similar local $\chi^2$ minima. This solution is activated when $\sigma_{bulge} + \sigma_{error} < \sigma_{disk}$, where $\sigma_{error}$ is the error in $\sigma$ from fitting a single component. Further details of the method are described in \citet{Oh20}. An example of the two-dimensional maps obtained for the kinematics and stellar populations of the bulge and disk components is shown in Figure~\ref{F31}. 

The mean age and metallicity of bulge and disk components are summarized in Table~\ref{Tab1}. The mean values are calculated from bins with bulge S/N, (S/N)$_{bulge} > 30$ and disk S/N, (S/N)$_{disk} > 20$, which are defined by (S/N)$_{bulge}$ = $\sqrt{B/T} \times$ (S/N)$_{continuum}$, and (S/N)$_{disk} = \sqrt{(1 - B/T)} \times$ (S/N)$_{continuum}$.

\begin{deluxetable*}{lcccccc}
\tablenum{1} \tablecolumns{7} \tablecaption{The measurements of luminosity-weighted mean age and metallicity for bulge and disk components for all our sample} \tablewidth{0pt}
\tablehead{Name & Stellar mass  & Local density  & Age$_{Bulge}$ &  [M/H]$_{Bulge}$ & Age$_{Disk}$ & [M/H]$_{Disk}$ \\
                & log(M$_{\star}$/M$_{\odot}$) & log$\rho_{M, 1Mpc}$ & (Gyr) &  & (Gyr) & }
\startdata
\hline
Lenticular galaxies \\
\hline
MCG-02-02-086 & 11.45      & 10.43        & 13.24                & $-0.15$                & 10.81                 & $-0.15$ \\
NGC1060       & 11.40      & 10.00        & 15.53                & $-0.02$                & 13.69                 & $-0.04$ \\
UGC10693      & 11.17      & \,\,\,9.88   & 11.83                & $-0.56$                & \,\,\,9.93            & $-0.57$ \\
NGC0507       & 11.16      & 10.47        & 14.58                & $-0.32$                & 12.72                 & $-0.35$ \\
UGC10905      & 11.07      & \,\,\,8.97   & \,\,\,9.04           & $-0.61$                & \,\,\,8.98            & $-0.54$ \\
NGC5513       & 10.89      & \,\,\,9.48   & 11.71                & $-0.69$                & \,\,\,9.96            & $-0.67$ \\
UGC02099      & 10.86      & \,\,\,8.65   & 4.99                 & $-1.12$                & \,\,\,6.96            & $-1.10$ \\
NGC0499       & 10.78      & 10.42        & 15.39                & $-0.03$                & 12.59                 & $-0.04$ \\
NGC7671       & 10.68      & \,\,\,9.30   & 14.13                & $-0.12$                & 14.49                 & $-0.23$ \\
NGC0364       & 10.65      & 10.56        & 12.68                & $-0.40$                & 12.60                 & $-0.46$ \\
PGC11179      & 10.64      & $-$          & 14.63                & $-0.13$                & 15.27                 & $-0.21$ \\
NGC0528       & 10.62      & 10.19        & 14.62                & $-0.24$                & 13.66                 & $-0.40$ \\
NGC0472       & 10.61      & \,\,\,9.93   & \,\,\,8.77           & $-0.80$                & \,\,\,7.73            & $-0.81$ \\
NGC7683       & 10.59      & \,\,\,9.90   & 11.76                & $-0.31$                & 12.59                 & $-0.31$ \\
NGC0515       & 10.58      & 10.68        & 13.53                & $-0.43$                & 14.39                 & $-0.41$ \\
NGC2476       & 10.58      & \,\,\,9.95   & 12.59                & $-0.96$                & 12.28                 & $-0.73$ \\
NGC0426       & 10.56      & 10.32        & 13.06                & $-0.24$                & 13.66                 & $-0.25$ \\
NGC1665       & 10.51      & \,\,\,8.85   & \,\,\,7.30           & $-0.59$                & \,\,\,6.64            & $-0.64$ \\
NGC0842       & 10.50      & 10.08        & 11.62                & $-0.92$                & 12.59                 & $-0.92$ \\
IC2341        & 10.47      & \,\,\,9.64   & 11.15                & $-1.01$                & $\;\;\;\,6.62^{\ast}$ & \,\,\,$-0.85^{\ast}$ \\
NGC7623       & 10.44      & 10.80        & 12.63                & $-0.38$                & \,\,\,15.51$^{\ast}$  & \,\,\,$-0.25^{\ast}$ \\
NGC7611       & 10.43      & 10.68        & 12.03                & $-0.41$                & \,\,\,8.77            & $-0.43$ \\
NGC0517       & 10.42      & 10.38        & 10.45                & $-0.62$                & \,\,\,9.06            & $-0.84$ \\
NGC2767       & 10.41      & 10.38        & 14.40                & $-0.03$                & 12.82                 & $-0.18$ \\
NGC6945       & 10.40      & \,\,\,9.77   & \,\,\,10.66$^{\ast}$ & \,\,\,$-0.54^{\ast}$   & $\;\;\;\,8.84^{\ast}$ & \,\,\,$-0.58^{\ast}$ \\
NGC6278       & 10.39      & \,\,\,9.80   & 12.90                & $-0.26$                & 15.78                 & $-0.39$ \\
NGC1281       & 10.37      & $-$          & 15.33                & \,\,\,\,\,$0.00$       & 11.93                 & $-0.08$ \\
NGC5473       & 10.35      & \,\,\,9.99   & 11.21                & $-0.37$                & \,\,\,9.43            & $-0.45$ \\
NGC6427       & 10.31      & \,\,\,9.82   & \,\,\,8.38           & $-0.57$                & 11.80                 & $-0.70$ \\
NGC2880       & 10.18      & \,\,\,9.55   & \,\,\,8.44           & $-0.27$                & \,\,\,6.44            & $-0.53$ \\
NGC5602       & 10.13      & \,\,\,9.40   & \,\,\,5.90           & $-0.60$                & \,\,\,6.45            & $-0.67$ \\
MCG-01-52-012 & 10.02      & \,\,\,9.53   & \,\,\,3.22           & $-1.25$                & \,\,\,4.05            & $-1.18$ \\
IC3586        & \,\,\,9.21 & \,\,\,9.76   & $\quad3.47^{\ast}$   & \,\,\,$-1.09^{\ast}$   & $\;\;\;\,0.95^{\ast}$ & \,\,\,$-1.33^{\ast}$ \\
IC3065        & \,\,\,9.00 & \,\,\,9.75   & $\quad0.75^{\ast}$   & \,\,\,$-1.20^{\ast}$   & \,\,\,2.84            & $-0.83$ \\
\hline
Passive spiral galaxies \\
\hline
UGC02018      & 10.73      & \,\,\,9.02   & \,\,\,9.84           & $-0.97$                & \,\,\,8.01           & $-1.05$ \\
UGC01271      & 10.54      & 10.38        & 12.59                & $-0.55$                & 12.39                & $-0.71$ \\
NGC7563       & 10.52      & 10.25        & 14.27                & $-0.20$                & \,\,\,14.45$^{\ast}$ & \,\,\,$-0.17^{\ast}$ \\
NGC2553       & 10.49      & 10.15        & 11.53                & $-0.58$                & 11.66                & $-0.69$ \\
NGC3300       & 10.40      & \,\,\,9.30   & 12.09                & $-0.35$                & 12.14                & $-0.50$ \\
NGC0495       & 10.38      & 10.51        & 12.19                & $-0.41$                & 14.10                & $-0.46$ \\
NGC5794       & 10.31      & 10.11        & 11.57                & $-0.60$                & \,\,\,15.85$^{\ast}$ & \,\,\,$-0.57^{\ast}$ \\
NGC5876       & 10.31      & \,\,\,9.78   & 11.53                & $-0.31$                & \,\,\,9.03           & $-0.40$ \\
NGC1666       & 10.23      & \,\,\,9.65   & \,\,\,9.69           & $-0.51$                & \,\,\,8.05           & $-0.71$ \\
\hline\hline
\enddata
\tablecomments{Column 1: the name of 34 lenticulars and nine passive spirals. Column 2: the stellar mass from NSA catalog. Column 3: Local density described in section 4.2. Column 4 and 5: the age and metallicity of bulge. Column 6 and 7: the age and metallicity of disk. The symbol of ‘$\ast$’ denotes the galaxies having insufficient bins ($< 50$) in the bulge and the disk.}
\label{Tab1}
\end{deluxetable*}

\section{Results}
\subsection{The age-metallicity relation of bulge and disk components}
\begin{figure*}
\centering
\includegraphics[width=17cm]{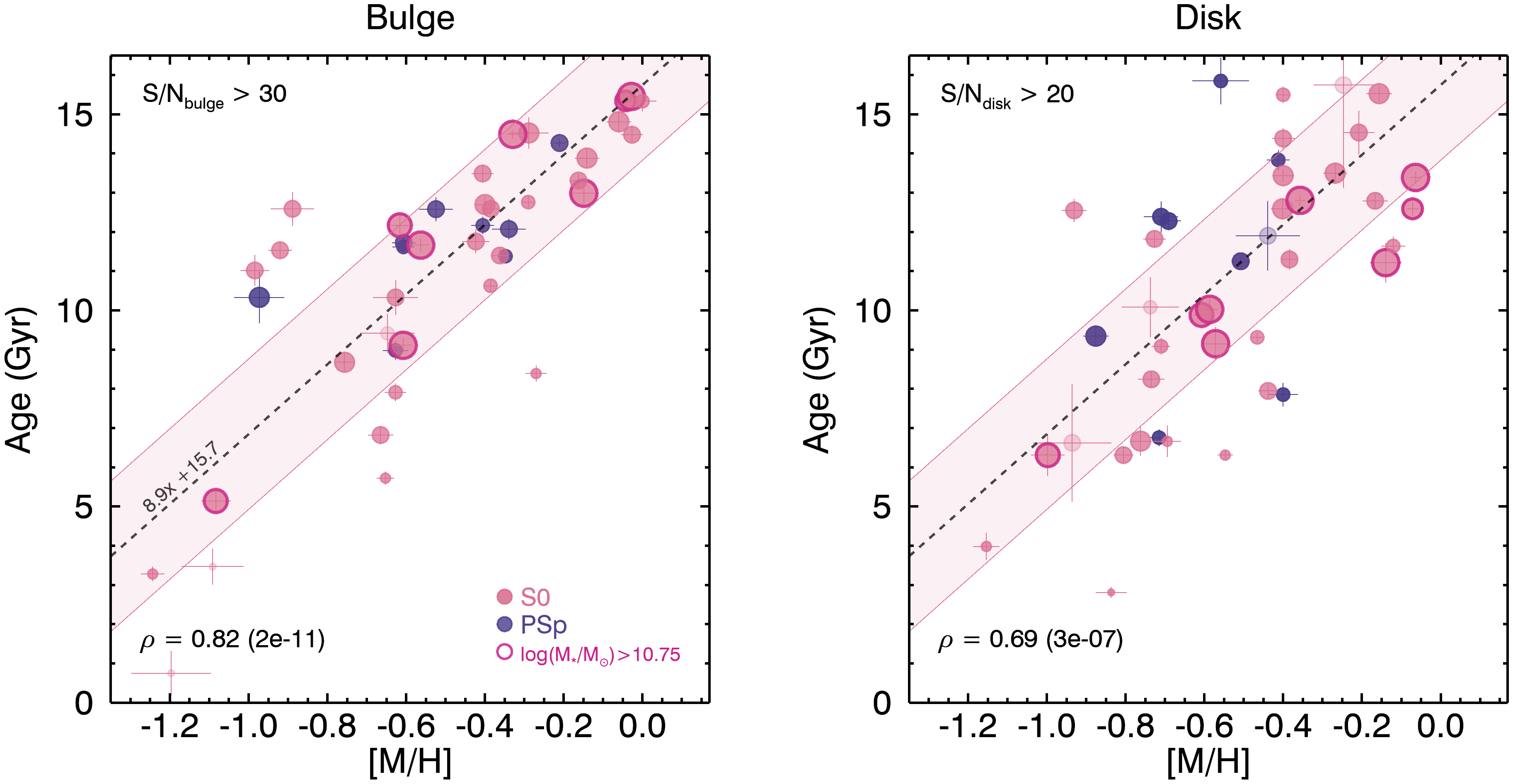}
\caption{Stellar ages vs.\ metallicities of the bulge (left panel) and disk (right panel) components for S0s (red circles) and passive spirals (dark blue circles). The size of a symbol is proportional to the stellar mass. The error bar indicates the standard deviation of the mean among the bins in each bulge or disk. The galaxies are divided into a high-mass subsample (log M$_{\star}$/M$_{\odot} > 10.75$; enclosed red open circles) and a low-mass subsample (log M$_{\star}$/M$_{\odot} \leq 10.75$). The fainter symbols indicate galaxies having insufficient bins in the bulge or the disk ($<50$). The dashed line in both panels is the linear fit for the bulges with more than $50$ bins and the shaded area shows the dispersion about this fit. The Spearman correlation coefficient and the significance of the null hypothesis are indicated in the bottom left of each panel.}
\label{F41}
\end{figure*}

\begin{figure}
\centering
\includegraphics[width=8cm]{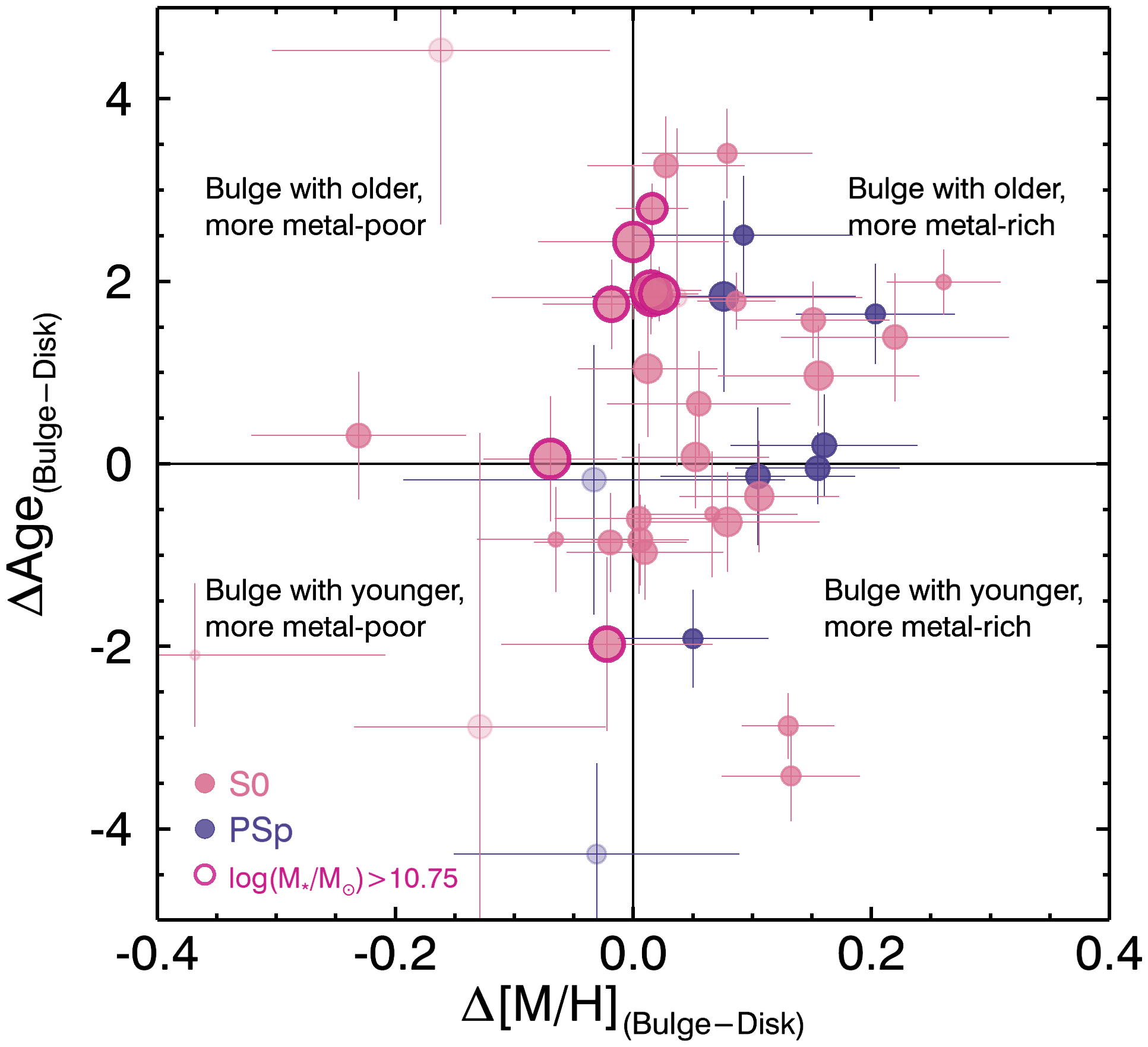}
\caption{The bulge-disk age difference $\Delta$Age$_{\rm (Bulge-Disk)}$ vs.\ the bulge-disk metallicity difference $\Delta$[M/H]$_{\rm (Bulge-Disk)}$ for individual galaxies. Symbols are the same as Figure~\ref{F41}.}
\label{F42}
\end{figure}

\begin{figure*}
\centering
\includegraphics[width=17cm]{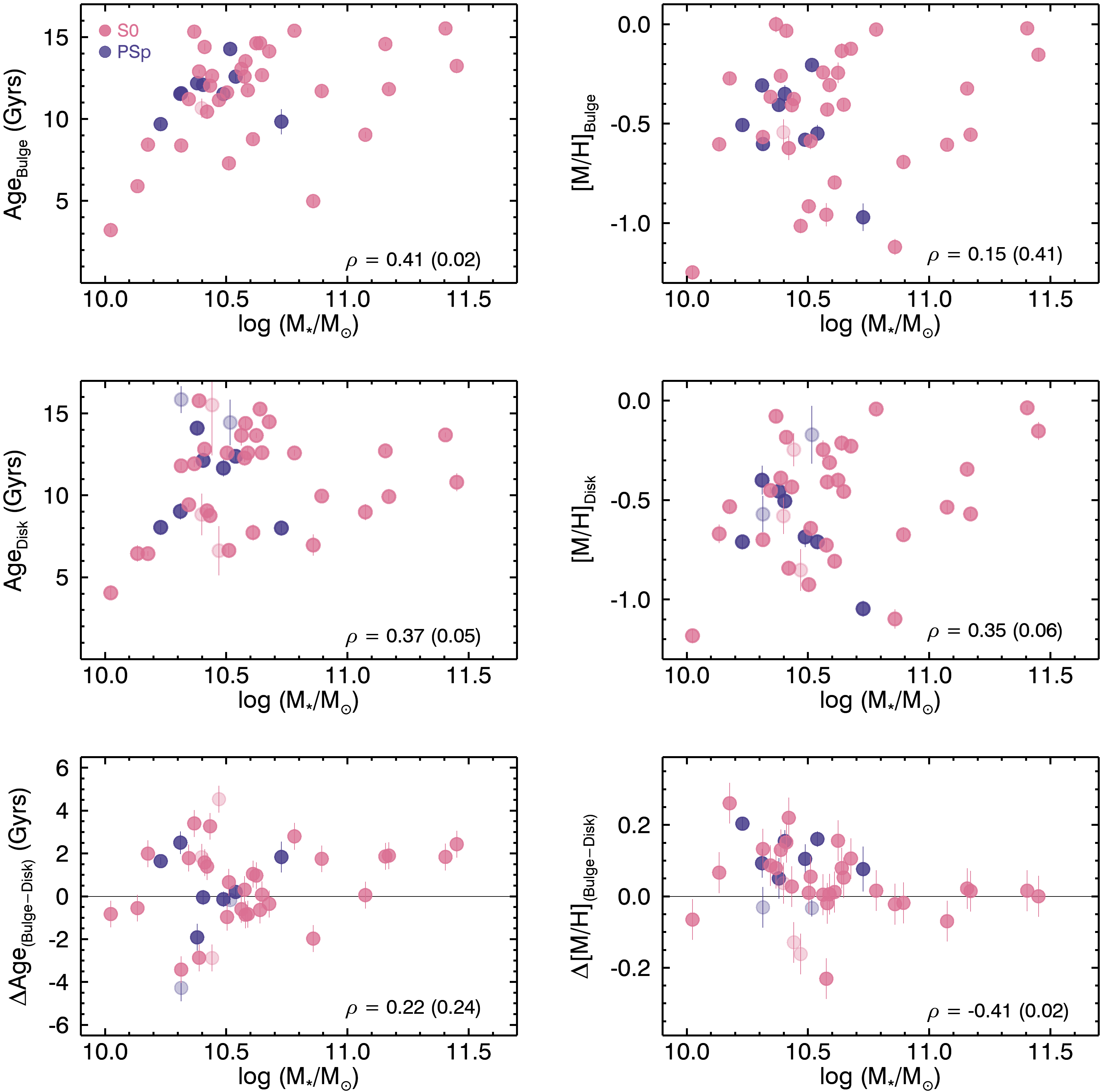}
\caption{The stellar ages (left panels) and metallicities (right panels) of bulges (top panels), disks (middle panels), and the difference between bulge and disk (bottom panels), all as functions of stellar mass.}
\label{F43}
\end{figure*}

Figure \ref{F41} presents the luminosity-weighted ages and metallicities of the bulge and disk components. We calculated the mean of total bulge and disk ages and metallicities only when the number of bins with sufficient S/N ($>30$ for bulges and $>20$ for disks) are $50$ or more. The bulges show a tight correlation, with a Spearman correlation coefficient ($\rho$) of $\sim 0.80$. The disks show a larger scatter than bulges, with $\rho \sim 0.66$, and are clearly offset from the robust least-squares linear fit to the bulges. We divide our S0s into a high-mass subsample (log M$_{\star}$/M$_{\odot} > 10.75$; $8$ S0s) and a low-mass subsample (log M$_{\star}$/M$_{\odot} \leq 10.75$; $26$ S0s, including two dS0); all passive spirals belong to the low-mass subsample. The bulges for low-mass S0s with low metallicity tend to be largely deviated from the fit. The disks of high-mass S0s are systematically offset from the robust least-squares linear fit to the bulges. The bulges of the passive spirals mostly lie well near the fit to the bulges. The disks of low-mass S0s and passive spirals widely scattered; unlike the disks of high-mass galaxies, the disks of some low-mass galaxies are as old as bulges. Although its faintness makes the signals low, the bulge and disk of two dwarf S0 (dS0) galaxy commonly have young ages.  
  
As seen in Figure~\ref{F42} and Figure~\ref{F43}, most bulges are older ($\sim 61\%$) and more metal-rich ($\sim 83\%$) than their associated disks. The metallicities of the bulges and disks are comparable for the high-mass S0s with standard deviation of $0.03$. The standard deviation of metallicity difference for the low-mass S0s is $0.14$. We can more clearly see this in comparing $\Delta$Age$_{(Bulge-Disk)}$ and $\Delta$[M/H]$_{(Bulge-Disk)}$ as a function of stellar mass in Figure \ref{F43}. Despite the large scatter, the age and metallicity of bulge and disk components appear to be correlated with their global stellar mass. In addition, galaxies with higher stellar masses have preferentially a larger $\Delta$Age$_{(Bulge-Disk)}$, and the deviation of $\Delta$Age$_{(Bulge-Disk)}$ is larger for galaxies with lower stellar masses. While $\Delta$[M/H]$_{(Bulge-Disk)}$ shows a negative correlations with stellar mass and is almost zero in the higher-mass galaxies. The passive spirals are also widely dispersed in $\Delta$Age$_{(Bulge-Disk)}$, and $\Delta$[M/H]$_{(Bulge-Disk)}$ is mostly positive. The results of Spearman's rank coefficient ($\rho$) and the significance of deviation (P$_{0}$) are summarized in Table~\ref{Tab2}.

\begin{deluxetable*}{lccccc}
\tablenum{2} \tablecolumns{6} \tablecaption{Summary of the Spearman correlation coefficient ($\rho$) and significance of deviation (P$_{0}$) for age and metallicity as a function of stellar mass and density} \tablewidth{0pt}
\tablehead{Y-axis & X-axis & total S0 & low-mass S0 & high-mass S0 \\
            & & $\rho$ (P$_{0}$) & $\rho$ (P$_{0}$) & $\rho$ (P$_{0}$)}
\startdata
Age$_{Bulge}$                    & log(M$_{\star}$/M$_{\odot}$) & 0.41 (0.02) & 0.51 (0.01) & 0.33 (0.42) \\
                                 & log$\rho_{M, 1Mpc}$          & 0.62 (0.0003) & 0.52 (0.02) & 0.79 (0.02) \\
\hline
Age$_{Disk}$                     & log(M$_{\star}$/M$_{\odot}$) & 0.37 (0.05) & 0.65 (0.00) & 0.36 (0.39)  \\
                                 & log$\rho_{M, 1Mpc}$          & 0.45 (0.02) & 0.38 (0.10) & 0.81 (0.01)  \\
\hline
$\Delta$Age$_{(Bulge - Disk)}$   & log(M$_{\star}$/M$_{\odot}$) & 0.46 (0.01) & 0.21 (0.35) & 0.05 (0.91)  \\
                                 & log$\rho_{M, 1Mpc}$          & 0.39 (0.05) & 0.24 (0.33) & 0.79 (0.02)  \\
\hline
$[$M/H]$_{Bulge}$                & log(M$_{\star}$/M$_{\odot}$) & 0.15 (0.41) & 0.18 (0.42) & 0.43 (0.30)  \\
                                 & log$\rho_{M, 1Mpc}$          & 0.39 (0.04) & 0.16 (0.47) & 0.74 (0.04)  \\
\hline
$[$M/H]$_{Disk}$                 & log(M$_{\star}$/M$_{\odot}$) & 0.35 (0.06) & 0.37 (0.09) & 0.38 (0.35) \\
                                 & log$\rho_{M, 1Mpc}$          & 0.43 (0.02) & 0.19 (0.42) & 0.69 (0.06) \\
\hline
$\Delta$[M/H]$_{(Bulge - Disk)}$ & log(M$_{\star}$/M$_{\odot}$) & -0.42 (0.02) & -0.19 (0.40) & 0.14 (0.74) \\
                                 & log$\rho_{M, 1Mpc}$          & 0.10 (0.64) & -0.10 (0.68) & 0.83 (0.01) \\
\hline\hline
\enddata
\tablecomments{The Spearman correlation coefficient ($\rho$) will be between a value of $-1$ and $+1$. $\rho = -1$ indicates a perfect negative correlation and $\rho = +1$ indicates a perfect positive correlation.}
\label{Tab2}
\end{deluxetable*}

\subsection{Environment}
\begin{figure*}
\centering
\includegraphics[width=17cm]{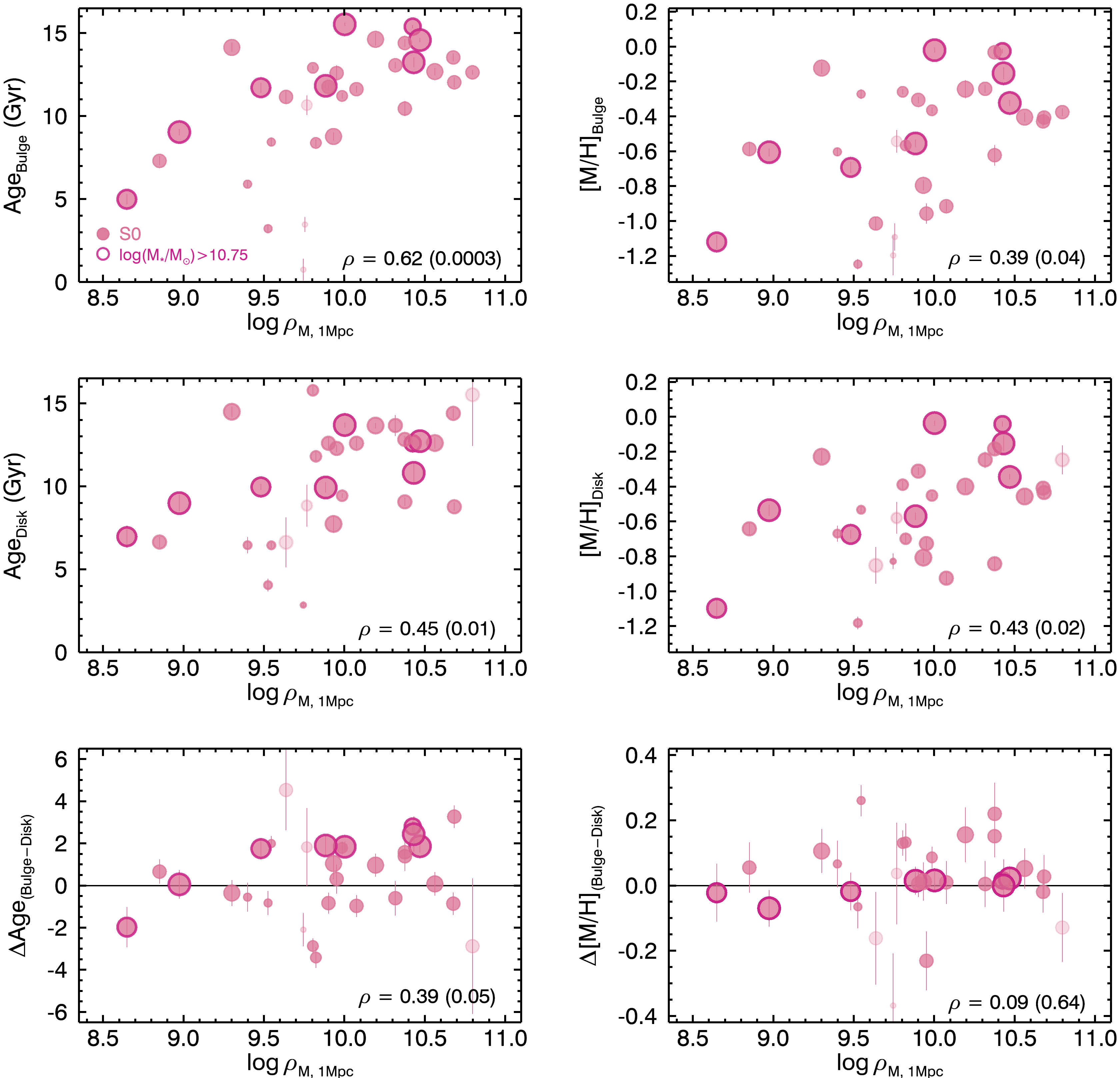}
\caption{The stellar ages (left panels) and metallicities (right panels) of bulges (top panels), disks (middle panels), and the differences between bulge and disk (bottom panels), as a function of local density; symbols are the same as Figure~\ref{F41}. The Spearman correlation coefficient and significance for all galaxies are shown at the bottom-right of the panels.}
\label{F46}
\end{figure*}
We investigate the environmental effects on the stellar populations of bulge and disk components. We calculate projected stellar mass density applying spline kernels (see Appendix~A2 in \citealt{Lee10}) by using the galaxies from the NASA-Sloan Atlas \citep{Bla11} within the velocity slice of $\pm500$\,km\,s$^{-1}$ and distance scale of $1$\,Mpc for each target galaxy. For completeness, we use the subsample of galaxies with mass $>10^{9.5}$\,M$_\odot$, but the mass of each target galaxy itself is not considered in calculating the mass density.

Figure~\ref{F46} presents the age and metallicity of each bulge and disk as a function of local density. It seems that both the stellar mass and environment do play a key role in the stellar populations of each bulge and disk. The ages of bulges and disks of high-mass S0s appear to increase with local density ($\rho = 0.79$ for bulges and $\rho = 0.81$ for disks). For low-mass S0s, the bulge ages also increase with local density ($\rho = 0.52$), but such a trend is unclear in the disk ages ($\rho = 0.38$). In addition, the $\Delta$Age$_{(Bulge-Disk)}$ tends to be constant with local density, both for the high-mass and low-mass S0s. The high-mass S0s have systematically higher $\Delta$Age$_{(Bulge-Disk)}$ than the low-mass S0s at given local density. 

The metallicity of bulges and disks of high-mass S0s increases with local density ($\rho = 0.74$ for bulges and $\rho = 0.69$ for disks), but low-mass S0s show no clear trend ($\rho = 0.16$ for bulges and $\rho = 0.19$ for disks). For the high-mass S0s, the metallicity in high-density environments appears to be larger than that in low-density environments both for bulges and disks, and $\Delta$[M/H]$_{(Bulge-Disk)}$ is zero on average, regardless of local density. The results of Spearman's rank coefficient ($\rho$) and the significance of deviation (P$_{0}$) are also summarized in Table~\ref{Tab2}.

\section{discussion}

Today, it is widely believed that there are two main pathways for the formation of S0s: (1)~fading of spirals to S0s and (2)~merger(s) and/or accretion. From a kinematical perspective, it is quite clear that slow-rotating S0s (stellar $v/\sigma < 0.5$) appear to be found preferentially in low-density environments, in accordance with the scenario that they are the descendants of mergers/accretion, while fast-rotating S0s dominate in denser environments, consistent with the spiral-fading scenario through the rapid consumption or removal of gas (\citealt{Dee20}; \citealt{Coc20}).

However, the detailed analysis of the decomposed stellar populations implies more complex stories than those revealed by kinematics only. From our results, bulges tend to be older in more massive host galaxies ($\rho = 0.41$), while the ages of the disks depend less on host galaxy mass ($\rho = 0.37$). This implies that a more massive galaxy forms its bulge earlier, whereas disk formation is barely dependent on the mass of the host galaxy and more extended in time. However, even at a fixed stellar mass, there is a large dispersion in stellar population properties, suggesting that either stochasticity or another physical factor other than stellar mass also influences a galaxy star-formation history.

Recently, the stellar populations in each component were analyzed through both line strength measurements and full spectral fitting to obtain estimates of the luminosity- and mass-weighted properties. \citet{Joh21}, where their sample of four S0s in the Centaurus Cluster and four isolated S0s observed with MUSE, revealed that the bulges are more metal-rich than the disks. They interpreted that the majority of the mass in these galaxies was built up early in the lifetime of the galaxy, with the bulges and disks forming from the same material through dissipational processes at high redshift. The younger stellar populations and asymmetric features seen in the field S0s may indicate that these galaxies have been affected more by minor mergers than the cluster galaxies. Using MaNGA sample, \citet{Tab19} showed bulges of S0s tend to be consistently more metal-rich than their disk counterparts, and while the ages of both bulge and disk are comparable, there is an interesting tail of younger and more metal-poor disks (see figure 2 of \citealt{Tab19}). On the other hand, \citet{Men19b} concluded that star formation only ever occurs in the disk of CALIFA S0s.

Since the luminosity-weighted age is sensitive to small fractions of recently generated stars, our analysis is more representative of recent star-formation histories than the mass-weighted age, which is representative of the average epoch when the bulk of the stars in a galaxy formed. From the analysis of 279 S0s from the MaNGA sample, \citet{Fra18} suggested that both mass and environment are important in the formation of S0s, but mass plays a more significant role than environment. However, although there is the caveat of small number statistics, our results about environments are robust. Bulges and disks tend to be older as the local density increases. The ages of disks increase for high-mass S0s but show a weak trend for low-mass S0s with local density. As a result, bulges are older than disks in high-density environments. This is consistent with the widely-believed scenario that S0s in dense environment were formed by major merger(s) or rapid collapse at an early epoch before falling into a higher-mass halo. Star formation in the early-formed bulges may have been quenched by some internal mechanisms, such as mass quenching (\citealt{Kau03}; \citealt{Wak12}) or AGN feedback (\citealt{Fab12} for a review; \citealt{Geo19}). The bulge quenching may not be due to environmental effects, because environmental effects must have also quenched the star formation in the disks, but disks are slightly younger than bulges at high density in our results. Thus, even after the bulge quenching, star formation in disks must have continued for some time. Alternatively, the disks may have re-formed for a time after the bulge was quenched. As time went on, those S0s may have moved to higher-density environments hostile to star formation in general. Eventually, star formation in the disks may have been also quenched by environmental effects, which explains why disks in high-density environments are older than those in low-density environments.

It is noted that high-mass S0s have slightly younger disks than bulges in high-density environments, but [M/H] is almost the same between bulge and disk. If the disks were created from surrounding matter ejected from early violent mergers, this result seems reasonable. The bulge and disk in an S0 may have similar metallicity because their sources are the same, and the later star formation may have made the disk younger without much altering its metallicity. Strangulation is a valid alternative explanation: the star-forming gas is contained by the surrounding medium and this forces higher recycling rates than possible in the field.

In low-density environments, even if high-mass S0s formed in the same way at an early epoch, they may be more easily rejuvenated by interactions with gas-rich neighboring galaxies (\citealt{San99}; \citealt{San08}), a process known to be more frequent in low-density environments with relatively low velocity dispersions (\citealt{Too72}; \citealt{Lav88}; \citealt{Byr90}). Subsequent new star formation may influence both age and metallicity in the host galaxies; whereas age always becomes younger in such interactions and subsequent star formation, the change in metallicity may largely depend on the properties of the interacting neighbors. In low-density environments, both bulges and disks seem to be influenced by gas accretion. Since gas stripping is not strong in low-density environments, they can keep forming stars using accreted gas and consequently contain relatively younger stars. In this scenario, therefore, $\Delta$Age and $\Delta$[M/H] are almost zero on average, regardless of mass. 

\begin{figure}
\centering
\includegraphics[width=8cm]{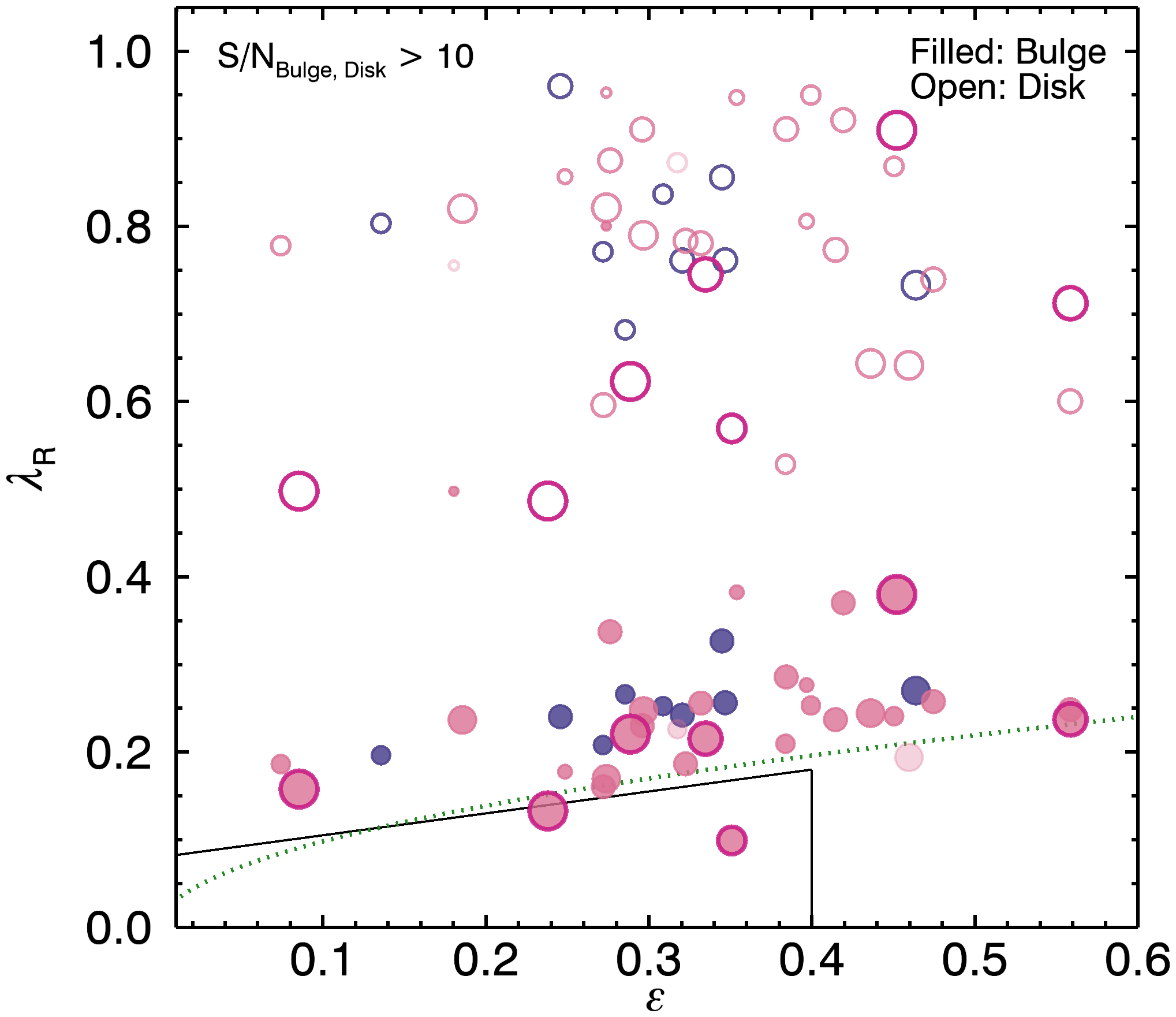}
\caption{The $\lambda_{R} - \epsilon$ diagram for bulges and disks; symbols are the same as Figure~\ref{F41}. We have measured $\lambda_R$ for bulges and disks following the equation (6) from \citet{Ems07}. For bulges, $\lambda_R$ is measured at R $<$ R$_e$. }
\label{F51}
\end{figure}

For low-mass S0s, both $\Delta$Age and $\Delta$[M/H] is almost constant with local density. At fixed local density, $\Delta$Age for low-mass S0s is systematically smaller than that for high-mass S0s; however, low-mass S0s at the highest density have larger $\Delta$Age than high-mass S0s at the lowest density. This result indicates that the impact on S0 formation of environment is more significant than that of stellar mass. The bulges and disks of low-mass S0s are preferentially fast rotators (Figure \ref{F51}). The relatively small age difference between bulge and disk and their fast rotation support the faded-spiral scenario (gas consumption by steady star formation) for the low-mass S0s, at least in low-density environments. On the other hand, the S0 galaxies with bulges much older than disks, found mostly in high-density environments, may have experienced environmental quenching (e.g.\ gas stripping) rather than steady fading. Unlike the bulges, the ages of the disks do not show a clear dependence on stellar mass. This may be because disks are more vulnerable to external influence than bulges. The large scatter in the age-metallicity relation of disks shows how complex their star formation histories are, compared to those of bulges.

Although it is not majority in our sample, there are $8$ out of $34$ S0s with younger age and more metal-rich than the corresponding disks, which are similar to the $13$ S0s of the Virgo cluster by decomposing long-slit spectroscopic data \citep{Joh14}. They are preferentially in lower-density environment in our sample, and most of them are low-mass S0s: \textbf{7 low-mass and one high-mass sample}. The $13$ S0s from \citet{Joh14} are preferentially in lower-density region or outskirt of the cluster with similar projected luminosity density to our sample, even though they are in the Virgo cluster. Their stellar masses (log M$_{\star}$/M$_{\odot}$) taken from NSA catalog also span a range of $9.2 - 10.6$ which is the same as our low-mass sample. \citet{Joh14} interpreted their results as evidence of late bulge starbursts activated by in-fallen gas, and our sample may share the same formation history with those S0s in the Virgo cluster. Since the Virgo cluster is an irregular and dynamically young cluster, further comparative studies with S0s in more regular and dynamically old clusters would be helpful to fully understand the evolution of S0s in various cluster environments.

\section{Conclusion}

We have analyzed the stellar populations of decomposed bulge and disk components for $34$ S0s using data from the CALIFA survey. The decomposed spectra of bulges and disks give several important constraints on the formation histories of S0 galaxies relating to their stellar mass and environment. In general, the stellar age-metallicity relation of bulges appears to be tight, whereas that of disks shows a negative offset from the bulge relation and has larger scatter. This implies that the star-formation history of the disks may be much more complex than that of the bulges. We found that age has more influence on the offset in the age-metallicity relation between bulge and disk than does metallicity. Most bulges are older than disks, while there is no systematic difference in metallicity.

Our results show that the formation histories of S0 galaxies depend on both their stellar mass and their environment. A more massive S0 galaxy tends to have a larger age difference between its bulge and disk. However, when stellar mass is controlled, an S0 galaxy in a higher-density environment tends to have a larger age difference. From these results, we suggest two scenarios of S0 formation depending on environment.\\

1. In dense environments, S0s were formed by major merger(s) or rapid collapse at an early epoch. The star formation in the early-formed bulges was quenched by internal mechanisms. After the bulge quenching, the ambient material ejected from the violent merger (re)formed disks, which continued their star formation for a long time (up to 5 Gyr from our results). As the S0s moved to higher-density environments over time, the star formation in disks was eventually also quenched, mainly by environmental effects such as gas-stripping.\\

2. In low-density environments, bulges were formed at an early epoch in a similar way to the bulges in high-density environments. However, unlike their high-density counterparts, they have been more easily rejuvenated by interactions with gas-rich neighbor galaxies. Both bulges and disks may have been influenced by gas accretion, and so the bulges and disks in low-density environment have very similar stellar populations. The major quenching mechanism in low-density environments may be simple fading through gas consumption by steady star formation.\\

In short, S0s in low-density environments may be faded spirals, while high-mass S0s in dense environments may have formed via dissipative merger(s), subsequent disk re-building, and finally environmental quenching. These scenarios are consistent with our overall results, but remain somewhat speculative and are too simple to explain all details. Thus, they need to be more rigorously tested and elaborated in further investigations. Most importantly, a much larger sample needs to be analyzed in a consistent method to ensure high statistical reliability.

\acknowledgments

We gratefully thank the anonymous referee for constructive comments that have significantly improved this manuscript. We are also grateful to Yujin Yang for helpful discussions. This study uses data provided by the Calar Alto Legacy Integral Field Area (CALIFA) survey (http://califa.caha.es/). Based on observations collected at the Centro Astron\'omico Hispano Alem\'an (CAHA) at Calar Alto, operated jointly by the Max-Planck-Institut f\"ur Astronomie and the Instituto de Astrof\'isica de Andaluc\'ia (CSIC). Parts of this research were conducted by the Australian Research Council Centre of Excellence for All Sky Astrophysics in 3 Dimensions (ASTRO 3D), through project number CE170100013. FDE acknowledges funding through the H2020 ERC Consolidator Grant 683184.

\end{document}